\documentclass[twocolumn,showkeys,superscriptaddress,showpacs,a4paper]{revtex4}
\usepackage{graphicx}
\usepackage{bm}
\usepackage{amsmath}
\usepackage{amsfonts}

\newcommand{\St}{{\mathrm{St}}}

\begin{document}

\title{Stochastic suspensions of heavy particles}

\author{J\'er\'emie Bec} \email{jeremie.bec@oca.eu}
\affiliation{Laboratoire Cassiop\'ee, Observatoire de la C\^ote
d'Azur, CNRS, Universit\'e de Nice Sophia-Antipolis, Bd.\ de
l'Observatoire, 06300 Nice, France}
\author{Massimo Cencini} \affiliation{SMC INFM-CNR c/o Dip.\ di Fisica
  Universit\`a di Roma ``La Sapienza'', Piazzale A.\ Moro 2, 00185
  Roma, Italy}\affiliation{CNR, Istituto dei Sistemi Complessi, Via
  dei Taurini 19, 00185 Roma, Italy}
\author{Rafaela Hillerbrand} \affiliation{The Future of Humanity
  Institute, University of Oxford, Suite 8, Littlegate House 16/17,
  St\,Ebbe's Street, Oxford, OX1\,1PT, United\,Kingdom}
\author{Konstantin Turitsyn} \affiliation{James Franck Institute,
University of Chicago, Chicago, IL 60637, USA } \affiliation{Landau
Institute for Theoretical Physics, Moscow, Kosygina 2, 119334, Russia}

\begin{abstract}
Turbulent suspensions of heavy particles in incompressible flows have
gained much attention in recent years. A large amount of work focused
on the impact that the inertia and the dissipative dynamics of the
particles have on their dynamical and statistical properties.
Substantial progress followed from the study of suspensions in model
flows which, although much simpler, reproduce most of the important
mechanisms observed in real turbulence. This paper presents recent
developments made on the relative motion of a pair of particles
suspended in time-uncorrelated and spatially self-similar Gaussian
flows.  This review is complemented by new results. By introducing a
time-dependent Stokes number, it is demonstrated that inertial
particle relative dispersion recovers asymptotically Richardson's
diffusion associated to simple tracers.  A perturbative
(homogeneization) technique is used in the small-Stokes-number
asymptotics and leads to interpreting first-order corrections to
tracer dynamics in terms of an effective drift.  This expansion
implies that the correlation dimension deficit behaves linearly as a
function of the Stokes number. The validity and the accuracy of this
prediction is confirmed by numerical simulations.
\end{abstract}



\maketitle

\section{\label{intro}Introduction}

\noindent
The current understanding of passive turbulent transport profited
significantly from studies of the advection by random fields. In
particular, flows belonging to the so-called \emph{Kraichnan ensemble}
--- i.\,e.\ spatially self-similar Gaussian velocity fields with no
time correlation --- which was first introduced in the late 1960's by
R.H.\ Kraichnan~\cite{k68}, led in the mid 1990's to a first
analytical description of anomalous scaling in turbulence
(see~\cite{fgv01} for a review). More recently, much work is devoted
to a generalization of this passive advection to heavy particles that,
conversely to tracers, do not follow the flow exactly but lag behind
it due to their inertia. The particle dynamics is thus dissipative
even if the carrier flow is incompressible.  This paper provides an
overview of several recent results on the dynamics of very heavy
particles suspended in random flows belonging to the Kraichnan
ensemble.

The recent shift of focus to the transport of heavy particles is
motivated by the fact that in many natural and industrial flows
finite-size and mass effects of the suspended particles cannot be
neglected. Important applications encompass rain
formation~\cite{pk96,pk97,ffs02} and suspensions of biological
organisms in the ocean~\cite{ro88,sf00,mopt02}. For practical
purposes, the formation of particle clusters due to inertia is of
central importance as the presence of such inhomogeneities
significantly enhances interactions between the suspended
particles. However, detailed and reliable predictions on collision or
reaction rates, which are crucial to many applications, are still
missing.

Two mechanisms compete in the formation of clusters. First, particles
much denser than the fluid are ejected from the eddies of the carrier
flow and concentrate in the strain-dominated regions~\cite{ef94}.
Second, the dissipative dynamics leads the particle trajectories to
converge onto a fractal, dynamically evolving
attractor~\cite{hc01,b03}. In many studies, a carrier velocity field
with no time correlation --- and thus no persistent structures --- is
used to isolate the latter effect. As interactions between three or
more particles are usually sub-dominant, most of the interesting
features of mono-disperse suspensions can be captured by focusing on
the relative motion of two particles separated by $\bm R$:
\begin{equation}
  \ddot{\bm R} = -\frac{1}{\tau} \left[ \dot{\bm R} - \delta\bm u(\bm
  R, t)\right]\,, \label{eqn:originalDyn}
\end{equation}
where dots denote time derivatives and $\tau$ the particle response
time. The fluid velocity difference $\delta\bm u$ is a Gaussian vector
field with correlation
\begin{equation}
  \left\langle \delta u^i(\bm r, t)\,\delta u^j(\bm r',
  t')\right\rangle = 2\, b^{\,ij}(\bm r - \bm r') \, \delta(t-t').
  \label{eqn:defcorrel}
\end{equation}
In order to model turbulent flows, the tensorial structure of the
spatial correlation $b^{ij}(\bm r)$ is chosen to ensure
incompressibility, isotropy and scale invariance, namely
\begin{equation}
  b^{ij}(\bm r) = D_1\,r^{2h}[(d-1+2h)\,\delta^{ij}-2h\,
  r^ir^j/r^2],
  \label{eqn:defdij}
\end{equation}
where $h$ relates to the H\"older exponent of the fluid velocity field
and $D_1$ measures the intensity of its fluctuations. In particular,
$h=1$ corresponds to a spatially differentiable velocity field,
mimicking the dissipative range of a turbulent flow, while $h<1$
models rough flows, as in the inertial range of turbulence. In this
paper we mostly focus on space dimensions $d=1$ and $d=2$; extensions
to higher dimensions are just sketched.

The above depicted model flow has the advantage that the particle
dynamics is a Markov process. In particular, Gaussianity and
$\delta$-correlation in time of the fluid velocity field imply that
the probability density $p(\bm r, \bm v, t |\bm r_0, \bm v_0, t_0)$ of
finding the particles at separation $\bm R(t) = \bm r$ and with
relative velocity $\dot{\bm R}(t) = \bm v$ at time $t$, when $\bm
R(t_0) = \bm r_0$ and $\dot{\bm R}(t_0) = \bm v_0$ is a solution of
the Fokker--Planck equation
\begin{equation}
  {\partial_t}p + \sum_i
  \left(\partial_{r}^i-\frac{1}{\tau}\partial_{v}^i\right) \!  \left(
  v^i p\right) - \sum_{i,j} \frac{b^{ij}(\bm r)}{\tau^2} \, \partial_v^i
  \partial_v^j p = 0,
  \label{eqn:fp}
\end{equation}
with the initial condition $p(\bm r, \bm v, t_0) \!=\! \delta(\bm r
\!-\!  \bm r_0)\,\delta(\bm v \!-\! \bm v_0)$.  To maintain a
statistical steady state, the Fokker--Planck equation~(\ref{eqn:fp})
as well as the stochastic differential
equation~(\ref{eqn:originalDyn}) should be supplemented by boundary
conditions, here chosen to be reflective at a given distance $L$.

For smooth flows ($h=1$), the intensity of inertia is generally
measured by the \emph{Stokes number} $\St$, defined as the ratio
between the particle response time $\tau$ and the fluid characteristic
time scale. For $\St\!\to\!0$, particles recover the incompressible
dynamics of tracers. In the opposite limit where $\St$ is very large,
inertia effects dominate and the dynamics approaches that of free
particles. In the above depicted model, the Stokes number is defined
by non-dimensionalizing $\tau$ by the typical fluid velocity gradient,
i.e.\ $\St \!=\!  D_1\tau$. Note that by rescaling the physical time
by $\tau$, it is straightforward to recognize that the dynamics
depends solely on $\St$.

Similarly it can be checked that in rough flows ($h\!<\!1$) --- with
an additional rescaling of the distances by a factor
$(D_1\tau)^{1/(2-2h)}$ --- the dynamics of a particle pair at a
distance $r$ only depends on the \emph{local Stokes number} $\St(r)
\!=\!  D_1\tau / r^{2(1-h)}$. This dimensionless quantity, first
introduced in~\cite{ffs03} and later used in~\cite{bch07}, is a
generalization of the Stokes number to cases in which the fluid
turnover times depend on the observation scale. At large scales,
$\St(r)\to 0$ and inertia becomes negligible. Particle dynamics thus
approaches that of tracers.  At small scales, $\St(r)\to \infty$ and
the particle and fluid motions decorrelate, so that the inertial
particles move ballistically. In both the large and small Stokes
number asymptotics, particles distribute uniformly in space, while
inhomogeneities are expected at intermediate values of $\St(r)$.

The paper is organized as follows.  In Section~\ref{piterbarg}, an
approach originally proposed in~\cite{p02} is used to reduce the
dynamics of the particle separation to a system of three stochastic
equations with additive noises. This formulation is useful for both
numerical and analytical purposes, particularly when studying the
statistical properties of particle pairs.  In Section~\ref{numerics},
we introduce the correlation dimension to quantify clustering as well
as the approaching rate which measures collisions. Numerical results
for these quantities are reported.  In Section~\ref{long-time-sep} we
introduce the notion of time-dependent Stokes number which makes
particularly transparent the interpretation of the behavior of the
long-time separation between particles. We show how Richardson
dispersion, as for tracers, is recovered in the long time
asymptotics. Section~\ref{oned} briefly summarizes some exact results
that can be obtained for the one-dimensional case.
Sections~\ref{smallstokes} and \ref{largestokes} are dedicated to the
small and large Stokes number asymptotics, respectively. In
particular, the former one presents an original perturbative approach
which turned out to predict, in agreement with numerical computations,
the behavior of the correlation dimension that characterizes particle
clusters. Finally, Section~\ref{turbu} encompasses conclusions, open
questions and discusses the relevance of the considered model for real
suspensions in turbulent flows.

\section{\label{piterbarg}Reduced dynamics for the two-point motion}

In this Section we focus on planar suspensions ($d=2$).  Following the
approach proposed in~\cite{p02} and with the notation $R= |\bm R| $,
the change of variables
\begin{eqnarray}
  \sigma_1 &=& (L/R)^{1+h} \bm R \cdot \dot{\bm R} / L^2,
 \label{eqn:DefReduced1}\\
   \sigma_2 &=& (L/R)^{1+h} |\bm R \wedge\dot{\bm R}| / L^2,
  \label{eqn:DefReduced2}\\
  \rho &=& (R/L)^{1-h}  
  \label{eqn:DefReduced3}
\end{eqnarray}
is introduced to reduce the original system of $2d\!=\!4$ stochastic
equations to the following one of only three equations
\begin{eqnarray}
  &&\dot{\sigma}_1 =  -\sigma_1/\tau - \left
  [{h\sigma_1^2-\sigma_2^2} \right]/ {\rho} + \sqrt{C}\,\eta_1,
  \label{eq:reduced1}\\
  &&\dot{\sigma}_2 = -\sigma_2/{\tau} - (1+h)
  {\sigma_1\sigma_2}/{\rho} + \sqrt{(1+2h)C}\,\eta_2,
  \label{eq:reduced2}\\
  &&\dot{\rho} = (1-h)\,\sigma_1,
  \label{eq:reduced3}
\end{eqnarray}
where $C \!=\!2 D_1/(\tau L^{1-h})^2$ and $\eta_i$ denote two
independent standard white noises. Reflective boundary conditions at
$R=L$ in physical space imply reflection at $\rho=1$.  Note that
$\sigma_1$ and $\sigma_2$ are proportional to the longitudinal and to
the transversal relative velocities between the two particles. In the
smooth case ($h=1$), one has $\rho=1$ and equations
(\ref{eq:reduced1}) and (\ref{eq:reduced2}) decouple from
(\ref{eq:reduced3}). The particle separation $R$ then evolves as
\begin{equation}
\dot{R} = \sigma_1(t) R\,.
\label{eq:reduced4}
\end{equation}

Besides this simple evolution and the reduction of the number of
variables from $2d$ to only three, the change of variables $\{{\bm
R},\dot{\bm R}\} \, \mapsto \,\{\rho,\sigma_1,\sigma_2\}$ has several
other advantages. For instance the noise, which is multiplicative in
the original dynamics~(\ref{eqn:originalDyn}), becomes additive in the
reduced system~(\ref{eq:reduced1})\,--\,(\ref{eq:reduced3}). However,
this simplification is counter-balanced by the presence of nonlinear
drift terms. Note that in dimensions higher than two, there is an
additional term $\propto 1/\sigma_2$, which is due to the It\^{o}
formula~\cite{dmow05,bch06}.

\begin{figure}[ht]
  \centerline{\includegraphics[width=0.55\textwidth]{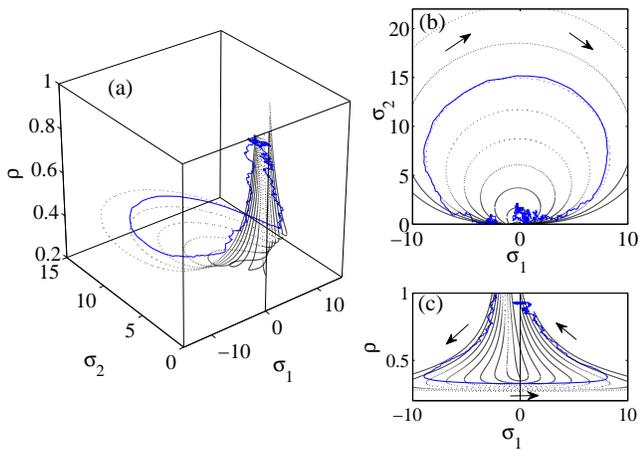}}
  \vspace{-10pt}
  \caption{\label{fig:drift2d_rough} Sketch of the reduced dynamics
    (\ref{eq:reduced1})\,--\,(\ref{eq:reduced3}) for $h=0.7$. The
    dotted lines represent the drift. The solid line depicts a random
    trajectory with $\St(L) = 1$. (a) full
    $(\sigma_1,\sigma_2,\rho)$-space, (b) projection on $\rho=0$
    plane, and (c) on the $\sigma_2=0$ plane. }
\end{figure}
Figure\,\ref{fig:drift2d_rough} sketches the deterministic drift and
shows a typical trajectory in the reduced space. This dynamics can be
qualitatively described as follows. The line
$\sigma_1\!=\!\sigma_2\!=\!0$ acts as a stable fixed line for the
drift. Hence a typical trajectory spends a long time diffusing around
it, until the noise realization becomes strong enough to let the
trajectory escape from the vicinity of this line. Whenever this
happens with a positive longitudinal relative velocity ($\sigma_1>0$),
the trajectory is pulled back to the stable line by the quadratic
terms in the drift. Conversely, if $\sigma_1\!<\!0$ and
$h\sigma_1^2\!+\!  \sigma_1\rho\! -\!\sigma_2^2\!  <\! 0$, the drift
pushes the trajectory towards larger negative values of
$\sigma_1$. Then the particles get closer to each other and $\rho$
decreases, until the quadratic terms in equations (\ref{eq:reduced1})
and (\ref{eq:reduced2}) become dominant. The trajectory then loops
back in the $(\sigma_1,\sigma_2)$-plane, approaching the stable line
from its right. It is during these loops that the inter-particle
distance $R$ becomes substantially small.  The loops thus provide the
main mechanisms for cluster formation.

\subsection*{Velocity statistics}

\begin{figure}[t!]
  \vspace{-5pt}
  \centerline{\includegraphics[width=0.55\textwidth]{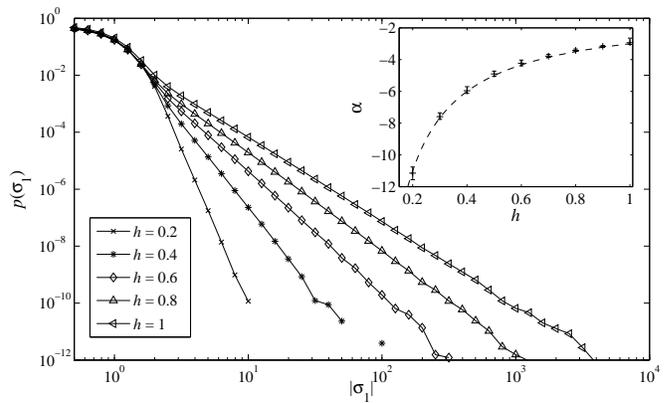}}
  \caption{\label{fig:tailpdfx} Log-log plot of the pdf of $\sigma_1$
    for $\St(L) \!= \!1$ for five values of the fluid H\"older
    exponent $h$. Power-law tails are always observed, $p(\sigma)
    \propto |\sigma|^{-\alpha}$. Inset: exponent $\alpha$ versus $h$;
    the dashed line is the theoretical prediction $\alpha\! =\!
    1\!+\!2/h$.}
\end{figure}

Numerical simulations show that the probability density function (pdf)
of the longitudinal relative velocity $\sigma_1$ displays algebraic
tails at large positive and negative values (see
Fig.~\ref{fig:tailpdfx}). As will become clear in the sequel, these
power-law tails are a signature of the above-mentioned large
loops. Let us consider the cumulative probability $P^<(\sigma) = {\rm
Pr}\,(\sigma_1 < \sigma)$ for $\sigma\!\ll\!-1$. This quantity can be
estimated as the product of (i) the probability to start a
sufficiently large loop in the $(\sigma_1,\sigma_2)$-plane that
reaches values smaller than $\sigma$ and (ii) the fraction of time
spent by the trajectory at $\sigma_1 \!  <\! \sigma$. Within a
distance of the order of unity from the line $\sigma_1
\!=\!\sigma_2\!=\!0$, the quadratic terms in the drift are subdominant
and can be disregarded. Then $\sigma_1 $ and $\sigma_2$ can be
approximated by two independent Ornstein--Uhlenbeck
processes. Conversely, at sufficiently large distances from that line,
only the quadratic terms in the drift contribute and the noise is
negligible.

Within this simplified dynamics, a loop is initiated at a time $t_0$
for which $\sigma_1(t_0)\!<\! -1$ and $\sigma_2(t_0)\!\ll
\!|\sigma_1(t_0)|$.  Once these conditions are fulfilled, the
trajectory performs a loop in the $(\sigma_1,\sigma_2)$-plane and both
$|\sigma_1(t)|$ and $\sigma_2(t)$ become very large. The maximum
distance from the stable line, which gives an estimate of the loop
radius, is reached when $\sigma_2$ is of the order of
$|\sigma_1|$. Let $t^\ast$ denote the time when this happens, i.e.\
$\sigma_2(t^\ast)/|\sigma_1(t^\ast)| \,=\, \mathrm{O}(1)$. When
neglecting the noise, this condition leads to the following estimate
for the loop radius
\begin{equation}
  |\sigma_1(t^\ast)| \propto
  [\sigma_1(t_0)+\rho(t_0)/\tau]\,|\tau\sigma_1(t_0)|^{h}
  \,(\tau\sigma_2(t_0))^{-h}\,,
\label{eq:rad}
\end{equation}
see~\cite{bch07} for details. In order to reach velocity differences
such that $\sigma_1\!<\!\sigma\!\ll\! -\!1$, the radius of the loop
has to be larger than $|\sigma|$. From (\ref{eq:rad}) this implies
that $\sigma_2(t_0)$ has to be smaller than $|\sigma|^{-1/h}$. In
order to evaluate contribution (i), one has to estimate the
probability to have $\sigma_1(t_0)\lesssim -1$ and
$\sigma_2(t_0)<|\sigma|^{-1/h}$ from the dynamics in the vicinity of
the origin. Approximating the two velocity differences $\sigma_1$ and
$\sigma_2$ by independent Ornstein--Uhlenbeck processes close to the
line $\sigma_1\!=\!\sigma_2\!=\!0$, the first condition gives an
order-unity contribution, while the second has a probability
$\propto\!  |\sigma|^{-1/h}$. For estimating (ii), we neglect the
noise in the dynamics far from the stable line. The probability is
then given by the fraction of time spent at $\sigma_1<\sigma$ which is
proportional to $\sigma_2(t_0) \!\propto\! |\sigma|^{-1/h}$. Put
together, the two contributions yield $P^<(x)\propto |\sigma|^{-2/h}$
when $\sigma\ll-1$. Thus the negative tail of the pdf of $\sigma_1$
behaves as $\propto |\sigma|^{-\alpha}$, with $\alpha = 1+2/h$.

During the large loops, the trajectories equally reach large {\em
positive} values of $\sigma_1$ and of $\sigma_2$. Again the fraction
of time spent at both $\sigma_1$ and $\sigma_2$ larger than $\sigma
\gg 1$ can be estimated as $\sigma^{-1/h}$. Hence, the pdf of both
longitudinal $\sigma_1$ and transversal $\sigma_2$ velocity
differences have algebraic left and right tails with exponent
$\alpha$. Both tails are depicted in Fig.~\ref{fig:tailpdfx}, where
the inset shows that the numerical measurements are in good agreement
with the predicted value of $\alpha$. The relation between $\alpha$
and the H\"older exponent $h$ implies in particular that $\alpha=3$ in
the smooth case, while it increases with decreasing $h$.  Moreover, it
follows straightforwardly from
(\ref{eq:reduced1})\,--\,(\ref{eq:reduced3}) that during the loops
$\rho(t) \propto \rho(t_0)^h$ when $\rho(t_0)\ll 1$. Hence it becomes
less and less probable to reach smaller values of $\rho$ as $h$
decreases. In other words, particle clustering should be very strong
for smooth flows and becomes weaker when the flow roughness is
increased.  This prediction is confirmed by the numerical studies
presented in next Section.

Finally it should be pointed out that although the change of variables
(\ref{eqn:DefReduced1})\,--\,(\ref{eqn:DefReduced3}) can be applied
equally in three dimensions, the above analysis does not carry over to
higher dimensions. Firstly, as already pointed out, an additional
drift term arises. This It\^{o}-term renders a straightforward
derivation of an analytic solution for the deterministic drift
impossible. Secondly, for higher dimensions the fixed point of the
reduced dynamics is located far from the origin, see
\cite{bch06}. Hence the approximations made above for $d=2$ are not
applicable. Careful numerical studies are needed to understand whether
or not algebraic tails are also present in higher dimensions.

%
%
%

\section{\label{numerics} Correlation dimension and approaching rate}

Particle clustering is often quantified by the \textit{radial
  distribution function} $g(r)$, which is defined as the ratio between
the number of particles inside a thin shell of radius $r$ centered
on a given particle and the number which would be in this shell if
the particles were uniformly distributed. This quantity enters
models for the collision kernel~\cite{rc00}.
Following~\cite{hc01,bccm05,bch06,bch07}, we consider a different,
but related way to characterize particle clustering. Instead of the
radial distribution function we evaluate the \textit{correlation
  dimension} $\mathcal{D}_2$ of the set formed by the particles. This
dimension is widely used in dissipative dynamical system theory and
in fractal geometry (see, e.g., \cite{er85,pv87}).  It is defined as
the exponent of the power-law behavior at small scales of the
probability $P_2(r)$ of finding two particles at a distance $R\! <\!
  r$:
\begin{equation}
  \mathcal{D}_2 = \lim_{r\to0} d_2(r),\quad d_2(r) = \frac{{\rm
  d} \ln P_2(r)}{{\rm d} \ln r}\,,
  \label{eq:DefCorrDim}
\end{equation}
where the logarithmic derivative $d_2(r)$ is called the \emph{local
correlation dimension}.  $\mathcal{D}_2$ relates to the radial
distribution function via $\ln g(r) / \ln r \to \mathcal{D}_2 \!-\! d$
for $r\!\to\!0$. For uniformly distributed particles,
$\mathcal{D}_2\!=\!d$, so that $g(r) \!=\!  \mathrm{O}(1)$. On the
contrary, when particles cluster on a fractal set,
$\mathcal{D}_2\!<\!d$ and $g(r)$ diverges for $r\!\to\!0$. This was
also found numerically in~\cite{rc00}.

Depending on whether the carrier flow is spatially smooth
($h\!=\!1$) or rough ($h\!<\!1$), $\mathcal{D}_2$ and $d_2(r)$
behave differently. In the former case, random dynamical system
theory \cite{a03} suggests that within the $2\times d$
position-velocity phase space, particles converge onto a
multifractal set with correlation dimension
$0<\overline{\mathcal{D}}_2<2d$. Here $\overline{\mathcal{D}}_2$
denotes the correlation dimension in the full phase space. It is
defined in complete analogy to $\mathcal{D}_2$ through the scaling
behavior of the probability $\overline{P}_2(r)$ to find two
particles at a distance less than $r$ in phase space:
\begin{equation}
  \overline{P}_2(r)\sim r^{\overline{\mathcal{D}}_2} \qquad
       {\mbox{for}} \qquad r\to 0\,.
  \label{eq:DefCorrDimPhaseSpace}
\end{equation}
The distance $r$ is now computed by using the phase-space Euclidean
norm $\sqrt{|\bm R|^2+|\bm V/D_1|^2}$; $\bm V$ is normalized by the
typical fluid velocity gradient $D_1$ for dimensional reasons. The
physical-space correlation dimension $\mathcal{D}_2$ is actually the
dimension of the projection of the set from the full phase space onto
the position space, and it is also expected to be fractal (see
Section~\ref{largestokes} for details on the relation between
$\overline{\mathcal{D}}_2$ and $\mathcal{D}_2$).  We focus in this
Section on quantifying clustering in position space and hence consider
only $\mathcal{D}_2$ and $d_2(r)$.

Balkovsky \textit{et al.}\ argued in~\cite{bff01} that particles do
not form fractal sets in non-smooth flows because the correlation
function of the particle density field should be a stretched
exponential.  Clustering and inhomogeneities are hence not quantified
by a fractal dimension but by the detailed scale dependence of
$d_2(r)$. However, as discussed in the Introduction, one expects the
statistical properties of two particles separated by a distance $r$ in
a flow with H\"older exponent $h$ to depend on the local Stokes number
$\St(r)\!=\!D_1\tau / r^{2(1-h)}$ only, which for smooth flows
degenerates to a scale independent number,
$\St(r)\!=\!\St\!=\!D_1\tau$. In rough flows, at scales small enough,
particles move ballistically and distribute homogeneously as the
Lagrangian motion is too fast for the particles to follow
($\St(r)\!\to\!\infty$ as $r\!\to\!0$) and hence $\mathcal{D}_2=d$ for
{\em all} particle response times $\tau$. However, information on the
inhomogeneities of the particle distribution at larger scales can
still be obtained through the scale-dependence of the local
correlation dimension $d_2(r)$ defined in (\ref{eq:DefCorrDim}).

\begin{figure}[b!]
  \centerline{\includegraphics[width=0.5\textwidth]{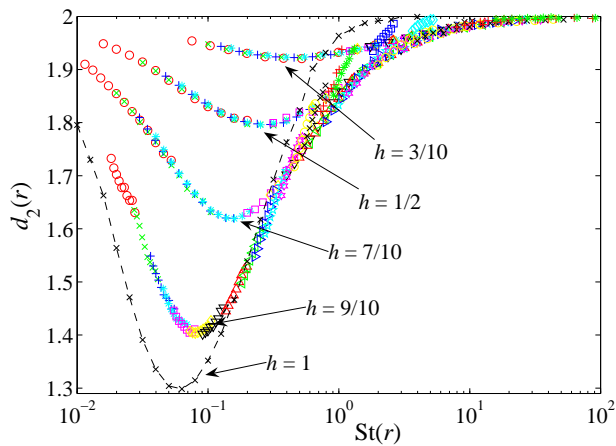}}
  \vspace{-10pt}
  \caption{\label{fig:collapsed2} Local correlation dimension $d_2(r)$
    versus the scale-dependent Stokes number
    $\St(r)\!=\!D_1\tau/r^{2(1-h)}$ for two-dimensional flows with
    different $h$. Symbols denote different particle response times
    $\tau$. For $h=1$, $\mathcal{D}_2=d_2(r\to 0)$ is
    displayed and $\St(r)=\St=D_1\tau$.}
\end{figure}
The relevance of the local Stokes number and of the local correlation
dimension is confirmed by numerical experiments of planar
suspensions. Simulations were performed by directly integrating the
reduced system described in previous Section. Figure
\ref{fig:collapsed2} shows $d_2(r)$ as a function of $\St(r)$ for
various values of $h$.  The curves obtained with different values of
the response time $\tau$ collapse onto the same $h$-dependent master
curve once the scale dependency is reabsorbed by using $\St(r)$. In
the plot, only scales far from the boundaries were considered, as
otherwise the self-similarity of the fluid flow is broken. The data
for $h=1$ estimate the limit of $d_2(r)$ as $r\!\to\!0$, and so
correspond to the value of the correlation dimension $\mathcal{D}_2$.
As anticipated in the previous Section, Fig.~\ref{fig:collapsed2} also
shows that clustering is weakening when the roughness of the fluid
velocity increases (i.e.\ when $h$ decreases). In particular,
$\min_{r}\{d_2(r)\}$ gets closer to $d$, i.e.\ particles approach the
uniform distribution as $h\to 0$. Finally notice that for $\St(r)\to
0$, i.e.\ at large scales in rough flows, $d_2(r)\!\to \!d$ as
well. This is due to the fact that at these scales the Lagrangian
motion becomes much slower than the relaxation time of the
particles. The particles thus recover the tracer limit and distribute
homogeneously.  As we will see in Section~\ref{smallstokes} the local
dimension $d_2(r)$ tends linearly to the space dimension $d$ when
$\St(r)\to0$ with a factor whose dependence on $h$ and $d$ can be
obtained analytically by perturbative methods.


The radial distribution function and hence the correlation dimension
give only partial information on the rate at which particles
collide. Indeed, in order to evaluate the collision rate,  one needs to
know not only the probability that the particles are close to each
other, but also their typical velocity difference. Here, following
\cite{bccm05}, we study the approaching rate $\kappa(r)$ defined as
the flux of particles that are separated by a distance less than $r$
and approach each other, i.e.\
\begin{equation}
  \label{eq:DefApproachingRate}
  \kappa(r) =\langle \dot{\bm R}\cdot {\bm R}/|{\bm
  R}|\Theta(-\dot{\bm R}\cdot {\bm R}/|{\bm R}|)\,\Theta(r - |\bm R|)
  \rangle \,,
\end{equation}
where $\Theta$ denotes the Heaviside function and the average is
defined on the Lagrangian trajectories.  As detailed in~\cite{bccm05},
$\kappa(r)$ is related to the binary collision rate in the framework
of the so-called \emph{ghost collision scheme} \cite{wwz98}. Within
this approach collision events are counted while allowing particles to
overlap instead of scattering. At small separations, $\kappa(r)$
behaves as a power law. This algebraic behavior allows defining a
\emph{local H\"older exponent} $\gamma (r)$ for the particle
velocities
\begin{equation}
  \gamma (r) = \frac{\ln \kappa(r)}{\ln r} - d_2(r)\,.
  \label{eq:gamma}
\end{equation}
In the above definition the contribution from clustering, accounted
for by the local correlation dimension $d_2(r)$, is removed.  The
local H\"older exponent $\gamma(r)$, similarly to $d_2(r)$, tends to a
finite limit $\Gamma$ as $r\to0$ which, for particles suspended in
a smooth flow ($h=1$), depends non-trivially on the Stokes number.

\begin{figure}[ht!]
  \vspace{-10pt}
  \centerline{\includegraphics[width=0.5\textwidth]{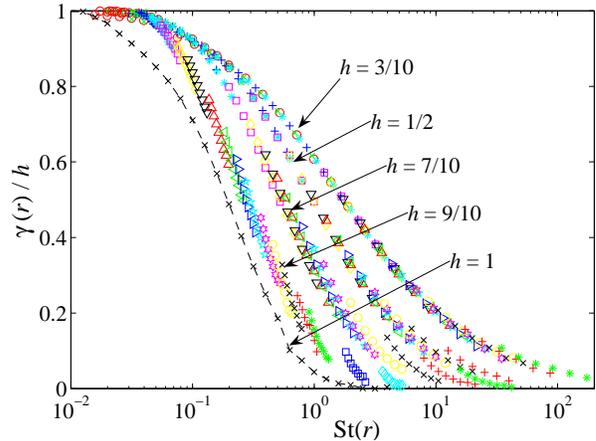}}
  \caption{Ratio between the local H\"older exponent $\gamma(r)$ of
    the particle velocity and that of the fluid $h$ versus
    $\St(r)$. The symbols in each curve refer to different values of
    the particle response time $\tau$. As in
    Fig.~\protect\ref{fig:collapsed2}, for $h=1$, the small scale
    limiting value $\Gamma$ is depicted. \label{fig:collapsegamma}}
\end{figure}

Figure~\ref{fig:collapsegamma} shows numerical estimations of
$\gamma(r)/h$ as a function of $\St(r)$ for various values of $h$.  In
the smooth case ($h\!  =\! 1$), the limit value $\Gamma$ decreases
from $\Gamma\!  =\!  1$ for $\St\!=\!0$, which corresponds to a
differentiable particle velocity field, to $\Gamma \!=\! 0$ for
$\St\!\to\!\infty$, which means that particles move with uncorrelated
velocities~\cite{bch06}. The fact that $\Gamma<1$ is due to the
contribution of caustics appearing in the particle velocity
field~\cite{mw04,wm05,mwdwl05,dmow05,bccm05} (see Sect.~\ref{oned} for
a discussion in $d=1$).  Similarly, in non-smooth flows $\gamma(r)$ is
asymptotically equal to the fluid H\"older exponent $h$ at large
scales ($\St(r)\!\to\! 0$), and approaches $0$ at very small scales
($\St(r)\!\to\!\infty$).  Therefore, all the relevant information is
entailed in the intermediate behavior of $\gamma(r)$. The latter
should only depend on the fluid H\"older exponent and on the local
Stokes number, as confirmed by the collapse observed in
Fig.~\ref{fig:collapsegamma}. Note that the transition from $\gamma(r)
= h$ to $\gamma(r)=0$ shifts towards larger values of the local Stokes
number and broadens as $h$ decreases.  The fact that $\gamma(r)=h$ for
$r\to \infty$ implies that the particles should asymptotically
experience Richardson diffusion just as tracers (see
Sect.~\ref{long-time-sep} for details). For comments on how the
findings reported in this Section translate to realistic turbulent
flows, we refer the reader to Section\,\ref{turbu}.

\section{\label{long-time-sep}Stretching rate and relative dispersion}

This Section is devoted to the study of the behavior of the distance
$R(t)$ between two particles at intermediate times $t$ such that $R(0)
\ll R(t) \ll L$. For convenience, we drop the reflective boundary
condition at $R=L$ and consider particles evolving in an unbounded
domain.

We first consider a differentiable fluid velocity field ($h=1$). In
this case, the time evolution of the distance $R(t)$ is given by
(\ref{eq:reduced4}), so that
\begin{equation}
  R(t) = R(0)\,\exp \left[{\int_0^{t} \sigma_1(t^\prime)\,\mathrm{d}t^\prime}\right]
  \label{eq:rfnX}
\end{equation}
and the particle separation can be measured by the \emph{stretching
rate} $\mu(t) \equiv (1/t)\ln [R(t) / R(0)]$. It is assumed that the
reduced dynamics (\ref{eq:reduced1})\,--\,(\ref{eq:reduced3}) is
ergodic. There is currently no rigorous proof of ergodicity.  However,
such an assumption relies on numerical evidence and on the following
phenomenological argument. The deterministic loops described in
Section~\ref{piterbarg} are randomly initiated by the near-origin
behavior of the system, providing a mechanism of rapid memory loss
that might ensure ergodicity.  With this assumption, the time averages
converge to ensemble averages, so that
\begin{equation}
  \mu(t) = \frac{1}{t} \int_0^{t} \sigma_1(t^\prime)\,\mathrm{d}t^\prime
    \to \langle \sigma_1 \rangle \quad\mbox{as } t\to\infty.
  \label{eq:lyap}
\end{equation}
In other words, the distance between particles asymptotically behaves
as $R(t) = R(0)\,\exp(t\lambda)$, where $\lambda = \langle \sigma_1
\rangle$ is a non-random quantity referred to as the \emph{Lyapunov
exponent}. A positive Lyapunov exponent implies that the particle
dynamics is chaotic~\cite{er85}.

\begin{figure}[t!]
  \centerline{\includegraphics[width=0.5\textwidth]{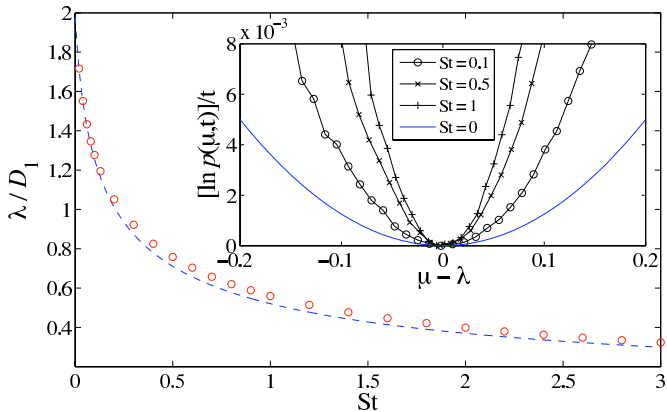}}
  \vspace{-10pt}
  \caption{\label{fig:lyapfit} Lyapunov exponent versus $\St$: the
  circles are the numerical measurements while the dashed line
  corresponds to Eq.~(\ref{eq:predlyap}). Inset: rate function $H$
  associated to the large deviations of the stretching rate $\mu$ for
  three values of $\St$; the solid line corresponds to $H$ for tracers
  for, whose analytic expression is known (see, e.g., \protect\cite{fgv01}).}
\end{figure}
Figure~\ref{fig:lyapfit} shows numerical measurements of the Lyapunov
exponent $\lambda$. The exponent remains positive for all values of
the Stokes number. This means in particular that particles suspended
in incompressible flow cannot experience \emph{strong clustering},
which consists in the convergence of all trajectories together to form
point clusters. This contrasts with the case of compressible flows
where, for suitable values of $\St$ and of the compressibility,
negative Lyapunov exponents are observed~\cite{mw04}. A first attempt
to derive an analytic expression for $\lambda(\St)$ was proposed by
Piterbarg~\cite{p02}. His approach is based on studying the Laplace
transform $\varphi(p)$ of the distribution of the complex random
variable $z=\sigma_1+i\sigma_2$, i.e.\ $\varphi(p,t) = \langle
\exp(-pz(t)) \rangle$ which satisfies
\begin{equation}
  \partial_t \varphi = -(p/\tau)\,\partial_p \varphi + p\,
  \partial_p^2 \varphi - (2D_1/\tau) p^2 \varphi.
  \label{eq:evollaplace}
\end{equation}
If $\varphi(p,t)$ reaches a steady state at large times, one can
infer an analytic expression for the asymptotic solution
$\varphi_\infty(p)$ by requiring that the right-hand side of
(\ref{eq:evollaplace}) vanishes. It is then straightforward to
deduce that the Lyapunov exponent satisfies $\lambda =
-\!\lim_{p\to0} \Re\{\partial_p\varphi_\infty\}$. This implies
\begin{equation}
  \lambda = - \frac{D_1}{2\St} \,\Re\!  \left\{1
  +\frac{\mathrm{A\!i}^\prime(x)}{\sqrt{x}\,\mathrm{A\!i}(x)}
  \right\}\!, \ x = (16\,\St)^{-2/3},
  \label{eq:predlyap}
\end{equation}
where $\mathrm{A\!i}$ and $\mathrm{A\!i}^\prime$ designate the Airy
function of the first kind and its derivative respectively. This
prediction is compared to the numerical measurements in
Fig.~\ref{fig:lyapfit}.  As stressed in~\cite{mwdwl05}, there is
evidence that the moments $\varphi(p,t)$ do not converge to a steady
state, but rather diverge at large times. This might explain the
discrepancies observed in Fig.~\ref{fig:lyapfit}. However, the
numerical precision is not high enough to test the presence of
corrections to the analytic expression~(\ref{eq:predlyap}).

At large but finite time $t$, the distance between the two particles
is measured by the \emph{stretching rate} $\mu(t) = (1/t)\ln
[R(t)/R(0)]$. This quantity becomes more and more sharply distributed
around the Lyapunov exponent $\lambda$ as $t$ increases. More
precisely, it obeys a large deviation principle and its pdf $p(\mu,
t)$ takes the asymptotic form (see, e.g.,~\cite{fgv01})
\begin{equation}
  \frac{1}{t} \ln p(\mu, t) \sim - H(\mu)\,,
  \label{eq:largdevstretch}
\end{equation}
where $H$ is a positive convex function attaining its minimum in
$\mu=\lambda$, in particular $H(\lambda)=0$.  The \emph{rate function}
$H$ measures the large fluctuations of $\mu$, which are important to
quantify particle clustering. Rate functions obtained from numerical
experiments are represented in Fig.~\ref{fig:lyapfit} for various
values of the Stokes number.  The function becomes less and less broad
when $\St$ increases, a phenomenon that can be quantified in the limit
$\St\to\infty$ as discussed in Section~\ref{largestokes}.  Note that
the same qualitative behavior is also observed for heavy particles
suspended in homogeneous isotropic flow~\cite{bbbcmt06}.


We now turn to the case of particles suspended in non-differentiable
flows ($h<1$). As we dropped the boundary condition, the initial
inter-particle distance $R(0)$ is the only relevant length scale.  By
using $R(0)$ instead of $L$ in the change of variables
(\ref{eqn:DefReduced1})\,--\,(\ref{eqn:DefReduced3}) the problem of
relative dispersion is expressed solely in terms of the the H\"older
exponent $h$ and of a time-dependent Stokes number which can be
defined in terms of the local Stokes number as $\St_t =
D_1\,\tau/[R(t)]^{2(1-h)}$. In particular, the evolution of $R(t)$
directly follows from the initial its value $\St_0$.  From the
evolution equation (\ref{eq:reduced3}) for the reduced separation
$\rho(t) = [R(t)/R(0)]^{1-h}$, we obtain
\begin{equation}
  \rho(t) = 1 + (1\!-\!h) \int_0^t \sigma_1(t^\prime) \,
  \mathrm{d}t^\prime \,,
  \label{eq:evolrho}
\end{equation}
where $\rho(\!\in\! [0,\infty))$ typically increases with time. The
time-dependent Stokes number $\St_t \!=\! D_1\tau/R^{2(1\!-\!h)} \!=\!
\St_0 / \rho^2$, which measures the effect of inertia when the
particles are at a distance $R(t)$, decreases with time. Hence,
conversely to the case of differentiable carrier flow, $\sigma_1$ is
not a stationary process and the integral in (\ref{eq:evolrho}) does
not tend to $t\langle\sigma_1\rangle$.

\begin{figure}[t!]
  \centerline{\includegraphics[width=0.5\textwidth]{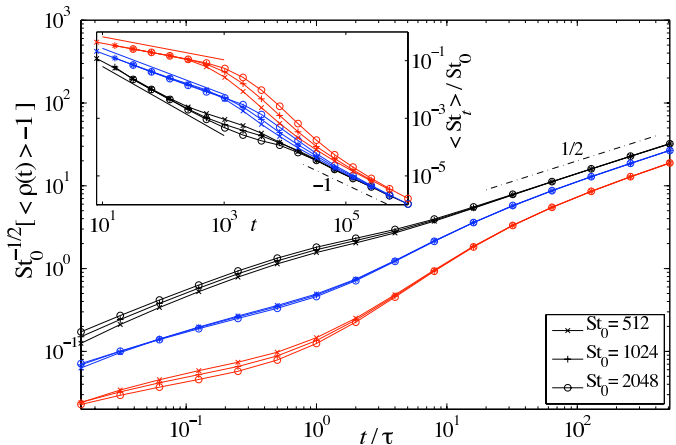}}
  \caption{\label{fig:reldisp}Time evolution of the average rescaled
    separation $\langle (\rho(t)-\rho(0)) \rangle$ for different
    initial Stokes numbers $\St_0$, and $h=0.4$, $0.6$, and $0.8$
    (from top to bottom).  Inset: long-time behavior of the
    time-dependent Stokes number $\St_t = D_1\tau/\rho^2(t)$ for
    different $\St_0$ and the same three values of $h$ (now from
    bottom to the top). The segments on the left indicate the slopes
    $-2/(1-h)$ corresponding to the regime of ballistic separation.}
\end{figure}

Hereafter, we confine the discussion to the case $\St_0\gg1$ because
it contains a richer physics than smaller $\St_0$.  As observed from
Fig.~\ref{fig:reldisp}, we can distinguish two regimes in the time
behavior of $\rho(t)$. At first the particle separation evolves
ballistically, i.e.\ $R(t)\propto t$, meaning that the time-dependent
Stokes number $\St_t$ decreases as $t^{-2/(1-h)}$ (see inset of
Fig.~\ref{fig:reldisp}) and reaches order-unity values for
$t\approx\tau$.  During this phase, the time growth of $\rho$ is
accelerated or slowed down and ultimately reaches a diffusive behavior
$\propto t^{1/2}$. This corresponds to the limit of tracers, which is
approached when $\St_t\ll1$. At this stage, the inter-particle
distance behaves as $R(t) \propto t^{1/2(1-h)}$ and, consequently, the
$\St_t$ decreases as $1/t$ (see Fig.~\ref{fig:reldisp}).

\begin{figure}[t!]
  \centerline{\includegraphics[width=0.5\textwidth]{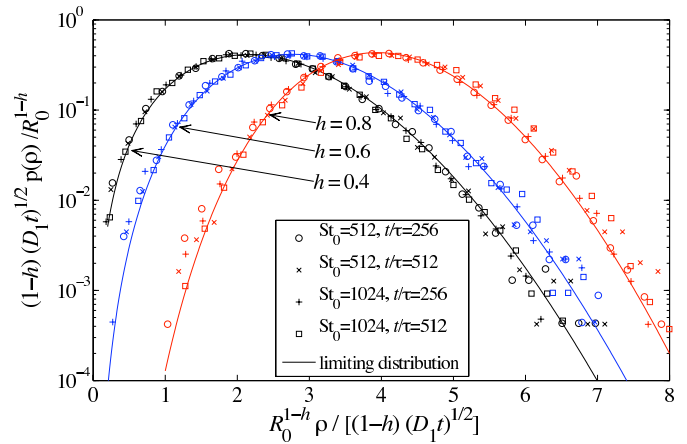}}
  \caption{\label{fig:fluctrho}Pdf of the rescaled separation
  $\rho(t)$ for various combinations of $\St_0$ and large times
  $t$. The solid lines represent the limiting distribution given by
  (\ref{eq:pdfrho}) with $A = 1/4$.}
\end{figure}

The convergence to tracer diffusion in the limit of large distances
$R$ gives an original way to interpret Richardson's law for
delta-correlated velocity fields in terms of the asymptotic behavior
of the reduced variables
(\ref{eqn:DefReduced1})\,--\,(\ref{eqn:DefReduced3}). When $\rho$ is
large, the quadratic terms in the drift of equation
(\ref{eq:reduced1}) can be neglected and $\sigma_1$ behaves as an
Ornstein--Uhlenbeck process with response time $\tau$. However, when
$\sigma_1$ becomes of the order of $\rho/(h\tau)$, the quadratic terms
cease to be negligible and they push the trajectory back to
$\sigma_1>0$. This process happens on time scales that are of the
order of unity and thus much smaller than the time scales relevant for
large-scale dispersion. Hence the dynamics of $\sigma_1(t)$ can be
approximated as an Ornstein--Uhlenbeck process with reflective
boundary condition on $\sigma_1 = \rho/(h\tau)$. This implies that
$\rho$ has a diffusive behavior. More specifically, numerical
simulations (see Fig.~\ref{fig:fluctrho}) show that the pdf of $\rho$
behaves as
\begin{equation}
  p(\rho,t) \propto {\rho^{\nu}}{t^{-(\nu+1)/2}} \exp \left[
  -A{\rho^2}/{t}\right],
  \label{eq:pdfrho}
\end{equation}
where $\nu = (1+h)/(1-h)$ and $A$ is a positive constant.  At large
times and consequently large distances $\St_t \to 0$, the tracer limit
is fully recovered as confirmed by expressing the above relation in
terms of the physical distance $R = \rho^{1/(1-h)}$. Indeed it becomes
identical to the law that governs the separation of tracers in a
Kraichnan flow~\cite{gv00}. However, a direct derivation of
(\ref{eq:pdfrho}) in terms of the $\rho$ and $\bm\sigma$ dynamics is
still lacking.

\section{\label{oned} Exact results in one dimension}

A number of analytical results were derived for one-dimensional flows
\cite{d85,wm03,dftt07}. Although such flows are always compressible,
their study helps improving the intuition for the dynamics of inertial
particles in higher-dimensional random flows. In particular, several
results on caustic formation hold also in two-dimensional
(incompressible) flows because the typical velocity fluctuations,
which lead to caustic formation, are effectively one-dimensional.

Here, we focus on one-dimensional smooth flows, for which the
equations analogous to (\ref{eq:reduced1})\,--\,(\ref{eq:reduced3})
reduce to
\begin{eqnarray}\label{eq:sigma1d}
  \dot{\sigma} & = & -\sigma/\tau - \sigma^2 + \sqrt{C}\,\eta(t),\\
  \dot{R} & = & \sigma R\,,\label{eq:R1d}
\end{eqnarray}
where $\sigma = V/R$ and, as in
(\ref{eq:reduced1})\,--\,(\ref{eq:reduced3}), $C=2D_1/\tau^2$.  The
quadratic term in~(\ref{eq:sigma1d}) implies that $\sigma$ can escape
to $-\infty$ with a finite probability. These events are the
one-dimensional counterpart of the loops described in
Section~\ref{piterbarg} and correspond to the formation of caustics:
particle trajectories intersect with a finite relative velocity. Note
that the equation for $\sigma$ decouples from the equation for $R$, so
that it can be studied separately. Stationary statistics of $\sigma$
can be described by the pdf $P(\sigma)$ which obeys the
one-dimensional Fokker-Planck equation
\begin{equation}\label{eqn:fp1d}
 \left[\partial_{\sigma}\left(\sigma/\tau +\sigma^2\right) +
 (C/2)\,\partial_\sigma^2\right] P(\sigma)=0\,.
\end{equation}
This equation can be rewritten as $\partial_\sigma J(\sigma) = 0$,
where $J(\sigma)=(\sigma/\tau+\sigma^2) P(\sigma)+C P'(\sigma)/2$ is a
probability flux in the $\sigma$-space.  Equation~(\ref{eqn:fp1d}) is
supplied by the boundary conditions $J(+\infty) = J(-\infty)$, which
are required to resolve escapes to infinity and thus caustic
formations. Indeed such events correspond to particle crossings during
which $R\to0$ and $V$ remains finite, so that $\sigma = V/R$ changes
sign. Hence, all particles escaping to $\sigma=+\infty$ reappear at
$\sigma=-\infty$. The stationary solutions of Eq.~(\ref{eqn:fp1d})
satisfying such a boundary condition corresponds to a constant flux
$J$ and can be written as
\begin{equation}
 P(\sigma)\! = \!\frac{2|J|}{C} \mathrm{e}^{-2\, U(\sigma)/C}
   \int_{-\infty}^\sigma \!\!\!\!\!\mathrm{d}\sigma'
   \mathrm{e}^{2\, U(\sigma')/C}\!,
   \label{constantfluxsolution}
\end{equation}
where $U(\sigma) = \sigma^3/3 + \sigma^2/2\tau$. Note that as in two
dimensions, $P(\sigma)$ has power-law tails. The argument presented in
Section~\ref{piterbarg} can actually be straightforwardly applied with
the difference that there is no loop anymore but just escapes to
infinity occurring with a probability that is independent of
$\sigma$. This leads to $P(\sigma) \propto |\sigma|^{-2}$ for
$|\sigma|\to\infty$ (the exponent is actually $-(1+1/h)$ in the
general case of H\"older-continuous carrier flows).

Using the constant-flux solution (\ref{constantfluxsolution}), one can
derive the Lyapunov exponent $\lambda=\langle \sigma\rangle$. As shown
in \cite{wm03}, its value non-trivially depends on the Stokes
number. For $\St = D_1 \tau \ll 1 $, it is negative and behaves like
$\lambda \simeq -D_1$ while for $\St\gg 1$ it becomes positive and its
value is given by the asymptotic expression $\lambda \simeq D_1
\St^{-2/3} \sqrt{3}\, 12^{5/6} \Gamma(5/6)/(24\sqrt\pi)>0$ . There
exists a critical value of the Stokes number ($\approx 0.827$) for
which the Lyapunov exponent changes its sign. This phenomenon of
sign-changing has been dubbed \emph{path coalescence transition} by
Wilkinson and Mehlig in \cite{wm03}. It is closely related to the
aggregation-disorder transition discussed in \cite{d85}. The sign of
the Lyapunov exponent determines how the distance between two
initially close particles evolves with time. It turns out that the
answer depends on the particle size: small particles (with
small-enough Stokes numbers) tend to approach each other, while large
particles (with large Stokes numbers) get separated by the flow.

Another important phenomenon which was extensively studied within the
one-dimensional model is the formation of caustics. The average rate
of caustics formation is given by the absolute value of the
probability flux $J$. For large values of the Stokes number it can be
written as $|J| \simeq D_1 \St^{-2/3} \Gamma(5/6) 12^{5/6}/(8
\pi^{3/2})$, while for small Stokes it becomes exponentially small
$|J|\sim D_1 (2\pi\St)^{-1}\exp[-1/ (6 \St)]$. The formation of
caustics is a stochastic process, whose properties can be described by
the pdf of the caustic formation time $T$. In \cite{dftt07} it is
shown that for $\St \ll 1$ this pdf can be estimated as $P(T)
\!\propto\! \exp[-1/(6 \St)]$ for $\tau \ll T \ll \tau \exp [1/(6
\St)]$ and $P(T)\!\propto \!\exp\left[-w/(3 CT^3)\right]$, with $w\!
=\! \Gamma(1/4)^8/ 96 \pi^2$ ($\Gamma$ denoting the Gamma-function
here), for $T \ll \tau$. The exponential factor $\exp[-1/(6 \St)]$
which characterizes the small rate of caustic formations for $\St \ll
1$ can be easily explained if one formally considers
Eq.~(\ref{eq:sigma1d}) as a Langevin equation for a particle which is
driven by the thermal noise $\eta(t)$ and evolves in the potential
$U(\sigma)$. In this case, the rate of caustic formation is given by
the probability for the particle to tunnel through the potential
barrier in $U(\sigma)$. Such probability can be estimated as
$\exp[-1/(6 \St)]$. For large Stokes numbers, the barrier disappears
and the rate of caustic formation is not exponentially damped anymore.

%
\section{\label{smallstokes}Small Stokes number asymptotics}

This Section reports some asymptotic results related to the limit of
small particle inertia. The first part summarizes the approach
developed by Mehlig, Wilkinson, and collaborators for differentiable
flows ($h\!=\!1$).  In analogy to the WKB approximation in quantum
mechanics (see, e.g.~\cite{l03}), the authors construct perturbatively
the steady solution to the Fokker--Planck equation associated to the
reduced system (\ref{eq:reduced1})\,--\,(\ref{eq:reduced2}). In the
second part of this Section original results are reported where the
particle dynamics is approximated as the advection by a synthetic flow
comprising an effective compressible drift which accounts for
leading-order corrections due to particle inertia.

Mehlig and Wilkinson proposed in~\cite{mw04} (see also~\cite{wmod07})
to approach the limit of small Stokes numbers in terms of the
variables $x_1\!=\!(\tau/D_1)^{1/2} \sigma_1$ and
$x_2\!=\!(\tau/3D_1)^{1/2} \sigma_2$.  From
equations\,(\ref{eq:reduced1})--(\ref{eq:reduced2}), their time
evolution follows to satisfy
\begin{eqnarray}
  \dot{x}_1 &=& -x_1 - \varepsilon \left[x_1^2 -3x_2^2 \right]
  + \sqrt{2}\,\eta_1(s)\,, \label{eq:redsmst1}\\ \dot{x}_2 &=&
  -x_2 - 2\varepsilon x_1x_2 +
  \sqrt{2}\,\eta_2(s)\,,\label{eq:redsmst2}
\end{eqnarray}
where $\varepsilon = \sqrt{\St}$, dots denote derivatives with
respect to the rescaled time $s\!=\!t/\tau$ and $\eta_1$ and
$\eta_2$ are independent white noises. The evolution equations
(\ref{eq:redsmst1})\,--\,(\ref{eq:redsmst2}) can be written in
vectorial form, namely  $\dot{\bm x} = -\bm x + \varepsilon
\mathbf{V}(\bm x) + \sqrt{2}\,\bm\eta$, where $\bm x = (x_1,x_2)$,
$\bm\eta =(\eta_1,\eta_2)$ and $\mathbf{V}$ denotes the quadratic
drift. The steady-state probability density $p(\bm x)$ is a solution
to the stationary Fokker--Planck equation
\begin{equation}
   \nabla_{\bm x}^2 p + \nabla_{\bm x}\cdot (\bm x p) = \varepsilon
  \nabla_{\bm x}\cdot [\mathbf{V}(\bm x) p]\,.
\end{equation}
Next step consists in writing perturbatively the probability density
of $\bm x$ as $p(\bm x) = \exp (-|\bm x|^2/4)\, (Q_0+\varepsilon Q_1 +
\varepsilon^2 Q_2 + \cdots)$. The functions $Q_k$ satisfy the
recursion relation $\mathcal{H}_0 Q_{k+1} = \mathcal{H}_1 Q_k$, where
\begin{eqnarray}
  \mathcal{H}_0 &=& 1 + \nabla_{\bm x}^2 - |\bm x|^2/4 \,, \\
  \mathcal{H}_1 &=& \nabla_{\bm x}\cdot \mathbf{V}(\bm x) + \bm x\cdot
  \mathbf{V}(\bm x)/2\,.
\end{eqnarray}
The operator $\mathcal{H}_0$ is the Hamiltonian of an isotropic
two-dimensional quantum harmonic oscillator. This suggests introducing
creation and annihilation operators and to expand the functions $Q_k$
in terms of the eigenstates of the harmonic oscillator (see
\cite{mw04,wmod07} for details).

This approach yields a perturbative expansion of the Lyapunov exponent
\cite{mw04}
\begin{equation}
  \lambda = D_1 \langle x_1 \rangle / \varepsilon = 2D_1 \sum_{k\ge 0}
  a_k \varepsilon^{2k} = 2D_1 \sum_{k\ge 0} a_k \St^{k},
  \label{eq:serielyap}
\end{equation}
where the coefficients $a_k$ satisfy the recurrence relation
\begin{equation}
  a_{k+1} = 4(4-3k)a_{k} -2 \sum_{\ell=0}^{k} a_\ell a_{k-\ell}\,,
\end{equation}
with $a_0=1$.  For large $k$, these coefficients behave as $a_k \sim
(-12)^k k!$, so that the series (\ref{eq:serielyap}) diverges no
matter how small the value of $\varepsilon$ (and thus of $\St$).
Hence the sum representation of $\lambda$ makes sense as an
approximation only if truncated at an index $k_\star$ for which
$|a_k\St^k|$ attains its minimum. For small values of $\St$,
$k_\star\sim 1/(12\St)$ and the error of the asymptotic
approximation is of the order of the smallest term, namely $\sim
|a_{k_\star}\St^{k_\star}| \sim \exp[-1/(12\St)]$. This approach was
refined by Wilkinson \textit{et al.}~\cite{wmod07} adopting an
approach based on Pad\'{e}--Borel summation, which was  found to
yield satisfactory results.

The non-analyticity of $\lambda(\St)$ at $\St=0$ is interpreted in
\cite{mw04} as a drawback of the perturbative approach. Indeed the
quadratic terms in (\ref{eq:redsmst1})\,--\,(\ref{eq:redsmst2}) are
not negligible for all values of $x_1$ and $x_2$: When $|\bm x|$
becomes larger than $\varepsilon^{-1}$ they are actually dominant and
the trajectory performs a loop in the $\bm x$ (or $\bm \sigma$) plane
(see Section\ \ref{piterbarg}). When $\St = \varepsilon^2$ is small,
the probability to initiate such a loop is given by the tail of the
distribution governing scales $|\bm x|\ll \varepsilon^{-1}$, and is
hence $\propto \exp [ -1/(6\varepsilon^2)]$, which coincides with the
one-dimensional result discussed in previous Section, confirming the
relevance of $d\!=\!1$ physics to the formation of caustics in higher
dimension. Taking into account this correction due to \emph{caustics},
i.e.\ the contribution of events when the particles approach very
close to each other keeping a finite velocity difference, Mehlig and
Wilkinson proposed to write the Lyapunov exponent as
\begin{equation}
  \lambda/D_1 \sim B\,\St^{-1}\mathrm{e}^{-1/(6\St)} + 2
  \sum_{k=0}^{k_\star} a_k \St^{k},
  \label{eq:serielyap2}
\end{equation}
where $B$ is a positive constant. We finish this summary by stressing
that this approach equally applies to the case of compressible carrier
flows~\cite{mw04}, and was extended to three dimensions where it
yields a prediction on the $\St$-dependence of the three largest
Lyapunov exponents~\cite{wmod07}.

The above perturbative approach can be generalized to small particles
evolving in rough flows. For small (local) Stokes numbers, the
characteristic time scales of velocity evolution are much smaller
compared to the temporal scales associated to the dynamics of the
particle separation. Therefore, one can obtain the effective equation
for the evolution of particle separation by averaging over the fast
velocity difference variables.  The systematic mathematical strategy
of such an averaging was proposed in \cite{mtv01} in the context of
stochastic climate models. This strategy is closely related to the
Nakajima--Zwanzig technique which was developed to study similar
problems arising in damping theory \cite{n58,z73}. Applications of
this technique to the elimination of fast variables in Fokker-Planck
equations are discussed in \cite{risken,cs79}.  In this framework one
can derive an expansion for the Fokker-Planck type operator entering
into the equation for the slow-variable probability distribution
function. In our case, this leads to a closed equation for the pdf of
the particle separation $R$. This equation can be used to determine
the local correlation dimension $d_2(r)$ for $\St(r)\ll 1$. We present
here only the general idea and the main results; details of the
calculations will be reported elsewhere.

\begin{figure}[t!]
  \centerline{\includegraphics[width=0.5\textwidth]{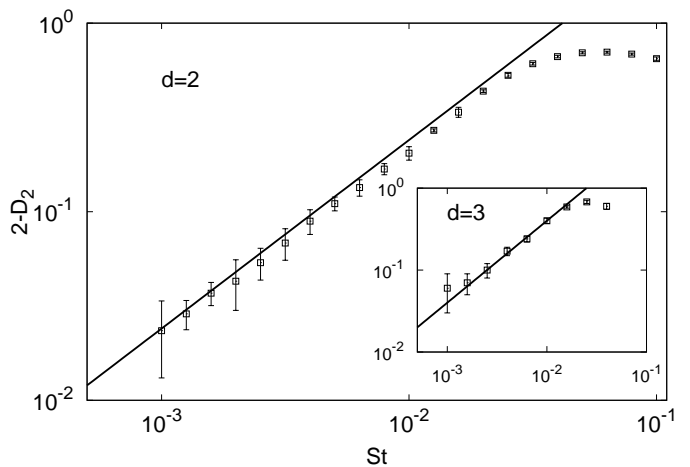}}
  \vspace{-10pt}
  \caption{\label{fig:d2smallstokes} Dimensional deficit
  $2-\mathcal{D}_2$ versus $\St$ in $d=2$ for smooth flows
  ($h=1$). Inset: same for $d=3$.  Points represent numerical results
  and the straight line corresponds to the perturbative predictions
  given by (\ref{eq:linearD2}) for $d=2$ and $3$ respectively.}
\end{figure}

To carry out the above-mentioned procedure the joint position-velocity
pdf $p({\bm r,\bm v})$ is approximated by
\begin{equation}\label{eq:ansatz}
  p({\bm r,\bm v}) \simeq p({\bm r}) P_{\bm r}({\bm v})
  +\tilde{p}({\bm r},{\bm v}),
\end{equation}
where $\tilde{p}({\bm r},{\bm v})$ denotes subleading terms which are
$\mathrm{O}(\St)$; $P_{\bm r}({\bm v})$ is the stationary distribution
associated to the fast velocity variables and satisfies the
Fokker-Planck equation
\begin{equation}\label{eq:L0}
 \hat L_0 \, P_{\bm r}({\bm v})\equiv-\left[\frac{1}{\tau}\partial_v^i
 v^i +\frac{b^{ij}({\bm r})}{\tau^2}\partial_v^i \partial_v^j
 \right]P_{\bm r}({\bm v}) = 0,
\end{equation}
with the normalization condition $\int \!\mathrm{d} {\bm v}\, P_{\bm
r}({\bm v}) = 1$.  Without loss of generality, it is assumed that the
subleading terms $\tilde{p}({\bm r},{\bm v})$ in the approximation
(\ref{eq:ansatz}) do not contribute to the normalization condition, so
that $\int \!\mathrm{d}{ \bm v}\, p({\bm r},{\bm v}) = p({\bm r})$.
The effective equation for $p({\bm r})$ can be derived by introducing
the expansion $p({\bm r}) = \sum_{k=0}^\infty \St^{k/2} p_k({\bm
r})$. This expansion, which enters the definition (\ref{eq:ansatz}),
is then substituted into (\ref{eqn:fp}) and all terms of the same
order in $\St$ are collected. Note, that the operator $\hat L_1 =
\partial_r^i v^i$ entering Eq.\ (\ref{eqn:fp}) is smaller than the
other operators by a factor $\St^{1/2}$. The chain of equations for
$p_k({\bf r})$ has a solvability condition that results in the
following effective equation for $p(\bm r)$:
\begin{equation} \label{eq:pert}
 \left(\hat M_1 + \hat M_2 + \cdots\right) p({\bm r}) = 0,
\end{equation}
where the operators $\hat M_k$ can be written as
\begin{equation}
 \hat M_{k}\, p({\bm r}) = \int\!\! \mathrm{d}{\bm v}\, \left(\hat L_1
 \hat L_0^{-1} \right)^{k} \hat L_1 \, p({\bm r})\, P_{\bm r}({\bm
 v}).
\end{equation}
$\hat L_0^{-1}$ denotes here the inverse of $\hat L_0$, i.e.\ the
Green function obtained from (\ref{eq:L0}) with the right-hand side
replaced by a $\delta$ function.  This operator is defined in such a
way that $\int\! \mathrm{d}{\bm v}\, \hat L_0^{-1} f({\bm v})\!  =\!
0$ for any function $f(\bm v)$ satisfying $\int\! \mathrm{d}{\bm v}\,
f({\bm v})\! =\! 0$.  One can check that the leading-order operator is
$\hat M_1 = \partial_r^i b^{ij}({\bm r})\partial_r^j$ which, as
expected, corresponds to turbulent diffusion. Indeed the dynamics of
tracers is recovered when $\St\to0$. The pdf $p({\bm r})$ which solves
the equation $\hat M_1 p({\bm r})\! =\! 0$ is simply the uniform
distribution. To measure particle clustering, which can be estimated
for instance by the local correlation dimension $d_2(r)$ (see
Section~\ref{numerics}), one has to calculate the next order
operators. It can be easily checked that all operators $\hat M_k$ of
even order $k$ are zero. The first non-vanishing correction to $\hat
M_1$ is thus given by the third order operator $\hat M_3$. When
interested in the stationary distribution only, the terms which enter
this operator and which are associated to transients can be
disregarded and one can write:
\begin{equation}
  \hat M_3 \,\bm\cdot = \partial_r^i[ V^i\,\bm\cdot\,], {\mbox{ with
  }} V^i\!=\!  -\frac{1}{2}\left(\partial_r^k\partial_r^l
  b^{ij}\right) \left(\partial_r^j b^{kl}\right).
\end{equation}
The operator $\hat M_3$ can be interpreted as an effective drift in
${\bm r}$-space and, for the Kraichnan model, represented as $V^i
\!=\! -\!2 (d^2\!-\!1)(d\!-\!2+4h)h^2 \St^2(r) r^i$. The functional
form of this drift implies that the first non-vanishing corrections to
the uniform distribution are proportional to $\St(r)$.  Indeed, for
isotropic flows one can look for a solution $p({\bm r})$, which
depends only on the modulus $r$ of its argument.  In this case Eq.\
(\ref{eq:pert}) becomes an ordinary differential equation of
Fokker-Planck type. Looking for a non-flux solution one readily
obtains the desired $p(r)$.  In rough flows ($h<1$), one has $\ln
p({\bm r})\sim [(d+1)(d-2+2h) h^2 /(1-h)]\, \St(r)$ and the local
correlation dimension behaves as
\begin{equation}
  d_2(r) \simeq d - \frac{2 d (d+1)(d-2+4h) h^2}{d-2+2h} \St(r).
\end{equation}
Note that the second term on the right-hand side of the above
expression disappears for $h\to 0$, confirming once again the finding
of the previous Sections about the decrease of clustering going from
smooth to rough flows.  For differentiable carrier flows ($h\,=\,1$),
the distribution has algebraic tails: $\ln p({\bm r})\!  \sim\!
-2(d+1)(d+2)\St \ln r$, and hence the correlation dimension behaves as
\begin{equation}
  \mathcal{D}_2 = d - 2(d+1)(d+2)\,\St + \mathrm{O}(\St^2).
  \label{eq:linearD2}
\end{equation}
The dimension deficit $d-\mathcal{D}_2$ is equal to $24 \St$ for
two-dimensional flows and to $d -\mathcal{D}_2 = 40 \St$ for
three-dimensional ones.  The latter result is in agreement with the
dimension deficit of the Lyapunov dimension reported by Wilkinson
\textit{et al.}\ in \cite{wmod07}. The above predictions on the
dimension deficit, for smooth flows, are in very good agreement with
numerical simulations in two and three dimensions, see
Fig.~\ref{fig:d2smallstokes}. We conclude this Section by noticing
that in time-correlated random smooth flows, as well as in developed
turbulence, the dimension deficit has been shown to be $\propto \St^2$
\cite{ffs02,b03,za03,fp04}. Therefore, including temporal correlations
seems to be crucial to reproduce the details of the small-Stokes
statistics of turbulent suspensions.

%
%
%
\section{\label{largestokes}Large Stokes number asymptotics}
Particles with huge inertia ($\St \gg 1$) take an infinite time to
relax to the velocity of the carrier fluid. They become therefore
uncorrelated with the underlying flow and evolve with ballistic
dynamics, moving freely and maintaining, almost unchanged, their
initial velocities.  This limit is particularly appealing for
deriving asymptotic theories~\cite{bch06}.  In this Section, we
focus on two aspects, namely the problem of the recovery of
homogeneous/uniform distribution for $\St \gg 1$ and the problem of
the asymptotic scaling for the statistics of the particle separation
and of the velocity differences.

\subsection{Saturation of the correlation dimension}
Ballistic particles injected homogeneously and uniformly remain
so~\cite{ss02}.  Hence for the correlation dimension associated with
their distribution (\ref{eq:DefCorrDim}) one has
$\mathcal{D}_2\!=\!d$.  This result follows directly from the
Fokker--Planck equation (\ref{eqn:fp}), which can be seen as an
advection-diffusion equation in phase space. The effective flow is
compressible because of the term $-\partial_v v/\tau$ but, in the
limit $St\to \infty$, it becomes negligible and the equation reduces
to diffusion plus advection by an incompressible flow. The resulting
stationary pdf is thus uniform in phase space and hence in its
projection in position space.  Moreover, as particle velocities and
fluid flow are uncorrelated and consequently the particles are not
correlated with each other, the exponent $\Gamma$ which characterizes
the small-scale behavior of the approaching rate (see
Section~\ref{numerics}) vanishes. Thus $\mathcal{D}_2\!\to\! d$ and
$\Gamma\!\to\!0$ for $\St\!\to\!\infty$.

This asymptotic regime can be achieved \textit{via} two possible
scenarios: (a) asymptotic convergence of $\mathcal{D}_2$ to $d$, and
(b) saturation of $\mathcal{D}_2$ to $d$ for Stokes numbers above a
critical value $\St^\dagger$. In what follows, we provide evidence for
(b), limiting the discussion to two-dimensional smooth flows.

Let us first discuss a phenomenological argument in favor of
saturation. As already noted in Section\,\ref{numerics}, their
dissipative dynamics yields the phase-space trajectories of the
particles to converge onto a random, dynamically evolving attractor,
which is typically characterized by a multifractal measure
\cite{er85,pv87}.  In our setting, this measure is the phase-space
correlation dimension defined in
equation\,(\ref{eq:DefCorrDimPhaseSpace}).  Ballistic motion for $\St
\gg 1$ corresponds to $\overline{\mathcal{D}}_2 \to 2d$, therefore a
critical Stokes number $\St^\dagger$ exists such that
$\overline{\mathcal{D}}_2(\St^\dagger)=d$. The particles' spatial
distribution is obtained by projecting the $(2\times d)$-dimensional
phase space onto the $d$-dimensional physical space. It is tempting to
apply a rigorous result on the projection of random fractal
sets~\cite{sy97,hk97} stating that for \textit{almost all}
projections, the correlation dimension of the projected set is related
to that of the unprojected one \textit{via} the relation
\begin{equation}
  \mathcal{D}_2=\min\{d,\overline{\mathcal{D}}_2\}\,.
  \label{eq:relbetweenD2s}
\end{equation}
Having $\overline{\mathcal{D}}_2(\St^\dagger)=d$ with the above
expression implies that $\mathcal{D}_2(\St)=d$ for all
$\St\ge\St^\dagger$. Unfortunately, there is \textit{a priori} no
reason for assuming some kind of isotropy in phase space which
justifies the validity of (\ref{eq:relbetweenD2s}). We thus proceed
numerically.

\begin{figure}[!t]
  \centerline{\includegraphics[width=.5\textwidth]{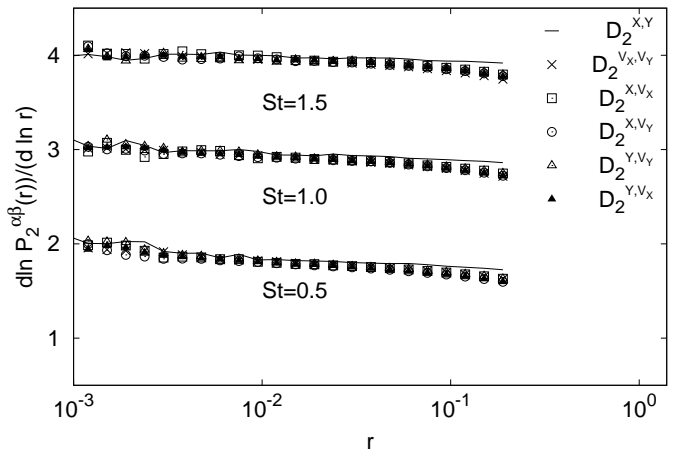}}
  \vspace{-10pt}
  \caption{\label{fig:projections} Logarithmic derivative $({\rm d}\ln
    P^{\alpha,\beta}_2(r))/({\rm d}\ln r)$ for different projections
    $\alpha,\beta$ for $\St=0.5$, $\St = 1$ (shifted up by a factor
    $1$), and $\St = 1.5$ (shifted up by a factor $2$). A small
    mismatch in the scaling range can observed for large $r$ (this is
    unavoidable as positions and velocities involve different
    scales).}
  \end{figure}

As Eq.~(\ref{eq:relbetweenD2s}) requires the isotropy of the set, we
have tested whether this applies to our case. The correlation
dimension of different two-dimensional projections was evaluated
through the computation of the probabilities $P^{\alpha,\beta}_2(r)$
of having two particles at a distance less than $r$ using the norm
$\Delta_{\alpha,\beta}^2 = \delta_{\alpha}^2+\delta_{\beta}^2$, with
$\alpha,\beta=X,Y,V_X/D_1,V_Y/D_1$, and $\delta_{\alpha}$ denoting the
coordinate-$\alpha$ separation between the two particles.  Note that
$\alpha=X$ and $\beta=Y$ corresponds to the spatial correlation
dimension discussed so far. Figure~\ref{fig:projections} shows the
logarithmic derivatives $({\rm d}\ln P^{\alpha,\beta}_2(r))/({\rm
d}\ln r)$ for various $\alpha,\beta$ and three different values $\St$.
All curves collapse within error-bars, confirming that the projection
is rather typical and thus strengthening the argument in favor of
saturation.
However, as can be seen in Fig.~\ref{fig:projections}, the logarithmic
derivatives on the different projections are curved, indicating
behaviors different from the expected power law.  It is therefore
difficult to decide whether or not the saturation occurs. As discussed
in~\cite{bclst04}, one can understand the curvature of the local
slopes with the presence of sub-dominant terms, e.g., with the
superposition of two power laws $P_2(r)\simeq A r^a+B r^b$. In our
case, one can expect that
\begin{equation}
P_2(r)= A r^{\overline{\mathcal{D}}_2}+B r^d\,,
\label{eq:two-powers}
\end{equation}
where $d$ and $\overline{\mathcal{D}}_2$ are the only dimensions
entering the problem~\cite{bch06}. For $\overline{\mathcal{D}}_2<d$,
the second power law can be interpret also as the contribution of
caustics~\cite{wm05,bccm05}: With non-zero probability, particles may
be very close to each other with quite different velocities, see
Section\,\ref{oned}. Once projected onto physical space, caustics
appear as spots of uncorrelated particles, and hence, the correlation
dimension is locally $\mathcal{D}_2=d$.  The validity of
(\ref{eq:two-powers}) as well as of the projection formula
(\ref{eq:relbetweenD2s}) was confirmed in Ref.~\cite{bch06}, .
\begin{figure}[ht]
  \centerline{\includegraphics[width=0.5\textwidth]{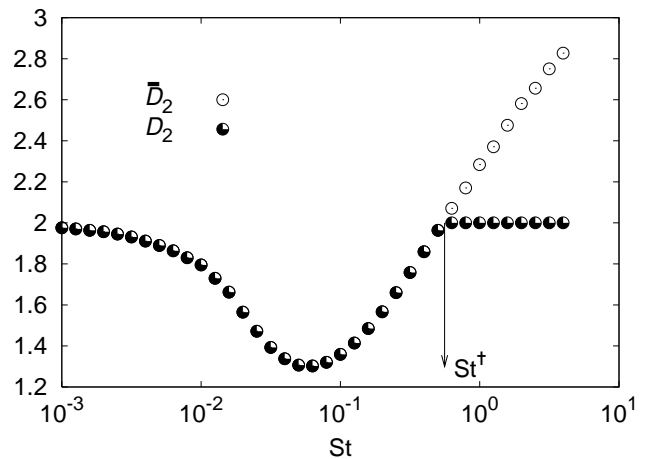}}
  \vspace{-10pt}
  \caption{Physical space $\mathcal{D}_2$, and phase-space
    $\overline{\mathcal{D}}_2$ correlation dimensions versus $\St$ as
    obtained by using (\ref{eq:two-powers}) for fitting the
    exponents. Errors are of the order of the size of the symbol. The
    arrow indicates the estimated location of
    $St^\dagger$.\label{fig:d2largestokes}}
\end{figure}

Figure~\ref{fig:d2largestokes} summarizes the results depicted
above. In particular, $\overline{\mathcal{D}}_2$ clearly displays a
crossover to values larger than $d$ for $St>\St^\dagger\approx 0.6$.
$\mathcal{D}_2$, once properly estimated by using
(\ref{eq:relbetweenD2s}), displays the saturation to $d=2$ above
$\St^\dagger$, at which the large Stokes asymptotics starts, at least
for the particle distribution.

Let us comment briefly on the implication of saturation on the
behavior of the approaching rate which, in the limit $\St\to\infty$,
is characterized by the exponent $\Gamma \to 0$. Similarly to
${\mathcal{D}}_2$, deviations of $\Gamma$ from its limiting value
cannot be determined by scaling arguments. Saturation of
${\mathcal{D}}_2$ would however affect $\Gamma$. This is related to
the dominant contribution of caustics which might imply also the
saturation of $\Gamma$ to $0$ for sufficiently large Stokes numbers.
Though numerical experiments confirm this scenario~\cite{bch06},
saturation cannot be studied with as much detail as for
${\mathcal{D}}_2$. At present, there is no simple phenomenological
argument for the subleading terms as for $\mathcal{D}_2$.

\subsection{Scaling arguments}

The limit of large values of the Stokes number can be approached by
assuming $\tau\to\infty$ and keeping $C=2D_1/(\tau L^{1-h})^2$
constant. The dynamics (\ref{eq:reduced1})\,--\,(\ref{eq:reduced2})
for the relative velocity differences can then be approximated by
\begin{eqnarray}
\label{eq:LimitingDyn1}
 \dot{\sigma}_1 &\simeq& - \left( h \sigma_1^2 - \sigma_2^2
 \right)/\rho + \sqrt{C}\, \eta_1\;, \\
\label{eq:LimitingDyn2}
 \dot{\sigma}_2 &\simeq& - (h+1) {\sigma}_1 {\sigma}_2/\rho +
 \sqrt{(1+2h) C}\,\eta_2\;.
\end{eqnarray}
For a given exponent $h$, the limiting dynamics depends solely on
$C$ while --- after non-dimensionalizing time and relative
velocities by $\tau$ --- the general dynamics depends on $\St(L)$
only (see the Introduction).  This congruence, which was first used
in \cite{h05} for determining the large-$\St$ behavior of the
Lyapunov exponent, allows to derive scaling arguments of various
other quantities characterizing two-particle dynamics.

Let us detail this for the distribution of the longitudinal velocity
difference $\sigma_1$. It is clear from the above considerations that
for fixed $h$ and $\sigma_1\gg(1/\tau)$ the following relation holds
\begin{equation}
\tau\, \tilde{p}(\tau \sigma_1; \St) \simeq p (\sigma_1; C)\,.
\end{equation}
Differentiating with respect to $D_1$ and $\tau$ gives a necessary
condition for such a behavior: $p$ must satisfy
\begin{equation}
  p + \sigma_1 \partial_{\sigma_1} p + 3C\, \partial_C p = 0\,,
\end{equation}
which itself implies $p(\sigma_1;C) = C^{-1/3} f(C^{-1/3}\sigma_1)$,
so that
\begin{equation}
  p(\sigma_1) \simeq \St^{-1/3} \tau f(\St^{-1/3} \tau\sigma_1)\,
  \qquad\mbox{for } \St \gg 1.
  \label{eq:scalingsigma}
\end{equation}

\begin{figure}[h]
  \centerline{\includegraphics[width=0.5\textwidth]{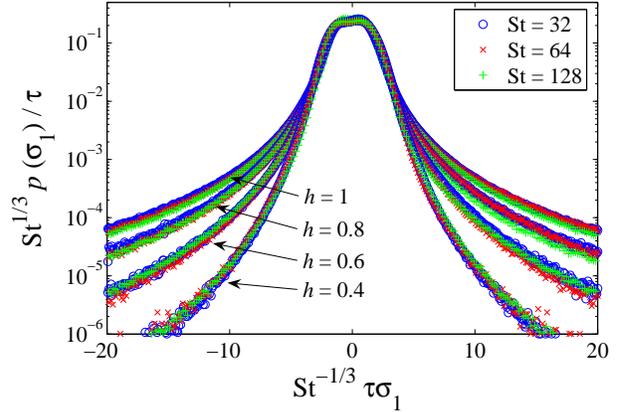}}
  \vspace{-12pt}
  \caption{\label{fig:collapse_pdfx} Pdf of the non-dimensional
    longitudinal velocity difference $\sigma_1$ at large values $\St$
    (symbols are for different values) for various values of $h$.}
\end{figure}
\begin{figure}[h]
  \centerline{\includegraphics[width=0.5\textwidth]{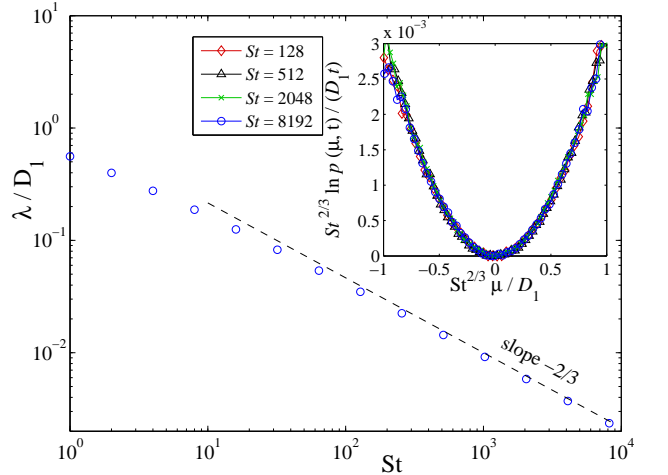}}
  \caption{\label{fig:lyap_largestokes}Lyapunov exponent $\lambda$
    versus $\St$. The dashed line is the asymptotic prediction
    (\ref{eq:ScalingLyap}). Inset: rate function $H(\mu)$ for various
    large values of $\St$.  }
\end{figure}

As shown in Fig.~\ref{fig:collapse_pdfx} this asymptotic scaling
behavior can be observed numerically.
As a consequence of (\ref{eq:scalingsigma}), for differentiable
carrier flows ($h=1$) the Lyapunov exponent $\lambda =
\langle\sigma_1\rangle$, which measures the asymptotic growth rate
of the inter-particle distance (see Section~\ref{long-time-sep}),
behaves as
\begin{equation}
\label{eq:ScalingLyap}
 \lambda \simeq c D_1 \St^{-2/3} \qquad\mbox{for }\St\gg1 \;,
\end{equation}
where $c$ is a parameter-independent positive constant. Note that the
original derivation~\cite{h05} of this law applies also to
compressible carrier flows, so the constant $c$ depends on the
compressibility of the fluid velocity field. It is shown
in~\cite{bch06} that this result also holds in three dimensions. Its
confirmation by numerical simulations is illustrated in
Fig.~\ref{fig:lyap_largestokes}.

The scaling argument described above can be carried forward to the
fluctuations of the stretching rate $\mu(t)=(1/t)\ln [R(t)/R(0)]$. As
we have seen in Section~\ref{long-time-sep}, for large times the
distribution of $\mu$ obeys the large deviation principle
(\ref{eq:largdevstretch}). It can be shown (see~\cite{bch06} for
details) that the associated rate function $H(\mu) =
\lim_{t\to\infty} (1/t) \ln p(\mu,t)$ satisfies
\begin{equation}
  H(\mu) \simeq D_1 \St^{-2/3} h(\St^{2/3}\mu/D_1) \qquad\mbox{for
  }\St\gg1.  \label{eq:ScalingStretch}
\end{equation}
This scaling is confirmed numerically (inset of
Fig.~\ref{fig:lyap_largestokes}).

We finally comment on how the stretching rate fluctuations change with
$\St$.  Taylor expansion of $H$ around its minimum together with the
scaling behavior (\ref{eq:ScalingStretch}) shows that the standard
deviation of the stretching rate is of the order of
$\St^{-1/3}/\sqrt{t}$. For a given time $t$, the stretching rate $\mu$
distributes more and more sharply around $\lambda$ when $\St$
increases. This behavior was anticipated by the numerical measurements
reported in Section~\ref{long-time-sep} and is observed in direct
numerical simulations of heavy particles in homogeneous isotropic
flows~\cite{bbbcmt06}.

%
%
%
\section{\label{turbu} Remarks and Conclusions}
Before concluding this paper the results discussed so far are
commented in the light of what is known about real turbulent
suspensions, which are relevant to most applications.  Let us start by
recalling the main features of turbulent flows.  Turbulence is a
multi-scale phenomenon~\cite{FrischBook} which spans length scales
ranging from a large (energy injection) scale $L$ to the very small
(dissipative) scale $\eta$, often called the Kolmogorov scale.  This
hierarchy of length scales is associated with a hierarchy of time
scales: from the large-scale eddy turnover time $\tau_L$ to the
Kolmogorov time $\tau_\eta$. Both ratios $L/\eta$ and
$\tau_L/\tau_\eta$ increase with the Reynolds number $\mathrm{Re}$ of
the turbulent flow.  Therefore, in general settings, no separation of
time scales can be invoked to simplify the motion of suspended
particles. However, in two circumstances simplifications are possible,
namely:\\
\noindent (i) For particles with a response time $\tau$ much greater
than $\tau_L$, the fluid velocity seen by the particle can be
approximated by a random flow belonging to the Kraichnan ensemble, as
discussed in this paper. Then a H\"older exponent $h=1$ or $h<1$ is
chosen to study the dissipative or inertial scales of turbulence,
respectively.\\
\noindent (ii) For intermediate response times
$\tau_\eta\ll\tau\ll\tau_L$, at least for single or two-particle
motions, the fluid velocity seen by the particles can be approximated
by an anisotropic generalization of the Kraichnan model~\cite{fouxon}.

In both asymptotics, the Kraichnan model and its generalization allow
for predictions on single- and two-particle properties, many of them
were discussed throughout this paper. In the following we discuss them
in the context of turbulent suspensions. We focus mostly on
two-particle properties at dissipative and inertial scales.

{\em Dissipative range\quad} At such small scales, particles form
(multi)fractal clusters, which can be quantitatively characterized by
the $\St$-dependence of the correlation dimension $\mathcal{D}_2$ or,
equivalently, of the dimensional deficit $d\!-\!\mathcal{D}_2$ (in
turbulence one can define $\St\! =\!  \tau/\tau_\eta$). Numerical
studies~\cite{hc01,bbclmt07} show that the qualitative
$\St$-dependence of $\mathcal{D}_2$ is similar to that observed in the
Kraichnan model.  Despite such similarities, it is likely that in
turbulence, ejection from vortical regions play, at least for small
$\St$, an important role~\cite{bbclmt07}. This can clearly not be
accounted for in Kraichnan flows, as $\delta$-correlated flows have no
persistent structures.  The absence of time correlations certainly
affects also the scaling behavior when $\St\!\ll\! 1$ of the dimension
deficit: while in turbulence~\cite{ffs02,fp04} and time-correlated
stochastic flows~\cite{b03,za03} it is observed that
$d\!-\!\mathcal{D}_2\propto \St^2$, we have shown here that the
behavior is linear in $\St$.  These discrepancies originate from the
fact that white-in-time carrier flows are valid approximations of
turbulence only for $\St\!\gg\!  1$.

Another question concerns the relative dispersion of a particle pair.
In the dissipative range, the velocity field is smooth, so that
particles separate exponentially with a rate given by the largest
Lyapunov exponent $\lambda$.  If $\tau\! \gg \!\tau_L$ the results
presented in previous Sections should apply, i.e.\ $\lambda
\!\propto\!\St^{-2/3}$. For $\tau_\eta \!\ll \!\tau\! \ll\!  \tau_L$,
the anisotropic generalization of the Kraichnan model predicts
$\lambda\!  \propto\!  \St^{-5/6}$~\cite{fouxon}. However, the
measurements of Lyapunov exponents made up to now (see
e.g.~\cite{bbbcmt06}) do not involve high-enough Stokes and Reynolds
numbers to test the validity of these predictions in turbulent flows

{\em Inertial range\quad} As shown in this paper, for rough
Kraichnan-type carrier flows, particles also form clusters which are
however not fractal as they were in the dissipative range. This seems
to be in qualitative agreement with the observations made in the
inertial range of turbulence: Inhomogeneities have been found in $2d$
turbulence in the inverse cascade regime~\cite{bdg04,cgv06} as well as
in $3d$ turbulence~\cite{bbclmt07,yg07}. However, while in the
Kraichnan case the particle distribution depends on the local Stokes
number $\St(r)$ only, this does not seem to be the case in turbulence,
at least for $\St(r)\ll1$ as studied in~\cite{bbclmt07} ( which in
turbulence is defined by $\St(r)\!=\!\tau/\tau_r$, $\tau_r$ being the
characteristic turbulent time scale associated to the scale $r$).  In
turbulent flows, for small values of $\St(r)$, a different rescaling
related to that of the acceleration (and hence pressure) field has
been found~\cite{bbclmt07}.  However such discrepancies do not
question the relevance of the Kraichnan model to turbulent flows as it
is expected to be a good approximation only for scales $r$ such that
$\tau_r\ll\tau$, i.e. $\St(r)\gg 1$.  Experiments or direct numerical
simulations with high $\mathrm{Re}$ and $\St$ are thus needed to
actually test the validity of the dynamical scaling in terms of
$\St(r)$ and to reproduce an equivalent of Fig.~\ref{fig:collapsed2}
for turbulent flows.  As far as particle separation is concerned, we
have seen in Section~\ref{long-time-sep} that at very long times, and
thus for separations $r$ such that $\tau\ll\tau_r$ one should expect
to observe Richardson dispersion.  For intermediate times at which the
separation is such that $\tau_\eta \ll \tau_r \ll \tau$, it is
predicted in~\cite{fouxon} that an intermediate asymptotic regime may
emerge with the typical particle separation $r$ growing as $t^9$,
i.e.\ much faster than Richardson diffusion.  On the numerical and
experimental side, we are not aware of any results on the relative
dispersion of two heavy particles in the inertial range. Testing the
above predictions can be probably done only in experiments where
$\mathrm{Re}$ can be very high.

In summary, this paper reviews most of current understanding of heavy
particle suspensions in Kraichnan-like stochastic flows. In
particular, we examined in details two-particle statistics both in
smooth and rough velocity fields.  Numerical simulations, validated by
analytics originally derived in this paper, show that particle
clustering is more efficient for smooth than rough flows, and can be
characterized in terms of the local Stokes number.  Detailed
predictions can be done in the very small and very large Stokes number
asymptotics. In the former we provided an analytical expression for
the dimensional deficit for any value of the fluid H\"older
exponent. More specifically, it is shown that the departure from a
uniform distribution is linear in the Stokes number, a result which is
confirmed by numerics.  As for the evolution of the relative
separation of particle pairs at small separations, a well-verified
asymptotic behavior for the Lyapunov exponent is discussed. At larger
scales, by converting the scale-dependent Stokes number into a
time-dependent one, we provided an original way to account for the
recovering of tracer-like Richardson diffusion. Finally, the relevance
of these results, together with other predictions obtained in recent
years from Kraichnan-like models of heavy particle suspensions, to
particles in turbulent flows has been discussed.

To conclude this work we suggest two different directions for further
investigations. First most of the predictions related to the
large-Stokes asymptotics lack numerical or experimental evidence in
fluid flows with high Reynolds numbers and particles with huge
inertia.  Second it is now definitely clear that an important
challenge for the near future is to understand whether or not some of
the techniques developed for suspensions in random time-uncorrelated
flows can be generalized/extended to time-correlated flows.  For
instance, a quantitative understanding of the small-Stokes-number
asymptotics in models that are closer to turbulence would be of great
interest to many applications. A first step in this direction has been
recently attempted in~\cite{musacchio}.

\begin{acknowledgments}
We acknowledge useful discussions with S.\ Musacchio and M.\
Wilkinson.  Part of this work was done while K.T.\ was visiting Lab.\
Cassiop\'{e}e in the framework of the ENS-Landau exchange program.
\end{acknowledgments}

\end{document}